\documentclass[%
 reprint,
superscriptaddress,
 amsmath,amssymb,
 aps,
]{revtex4-2}

\usepackage{comment}
\usepackage[caption=false]{subfig}
\usepackage{gensymb}
\usepackage{upgreek}
\usepackage{graphicx}% Include figure files
\usepackage{dcolumn}% Align table columns on decimal point
\usepackage{bm}% bold math
\usepackage{textcomp}

\begin{document}
\preprint{APS/123-QED}

\title{Terahertz-driven electron acceleration from subrelativistic to fully relativistic energies in stepped dielectric-lined waveguides}

\author{Filip J. Peczek}
\affiliation{The Cockcroft Institute, Sci-Tech Daresbury, Keckwick Lane, Daresbury, Warrington WA4 4AD, United Kingdom}
\affiliation{Department of Physics and Astronomy \& Photon Science Institute, The University of Manchester, Oxford Road, Manchester M13 9PL, United Kingdom}

\author{Laurence J. R. Nix}
\affiliation{The Cockcroft Institute, Sci-Tech Daresbury, Keckwick Lane, Daresbury, Warrington WA4 4AD, United Kingdom}
\affiliation{School of Engineering, Bailrigg, Lancaster LA1 4YW, United Kingdom}

\author{Graeme Burt}
\affiliation{The Cockcroft Institute, Sci-Tech Daresbury, Keckwick Lane, Daresbury, Warrington WA4 4AD, United Kingdom}
\affiliation{School of Engineering, Bailrigg, Lancaster LA1 4YW, United Kingdom}

\author{Rosa Letizia}
\affiliation{The Cockcroft Institute, Sci-Tech Daresbury, Keckwick Lane, Daresbury, Warrington WA4 4AD, United Kingdom}
\affiliation{School of Engineering, Bailrigg, Lancaster LA1 4YW, United Kingdom}

\author{Joseph T. Bradbury}
\affiliation{The Cockcroft Institute, Sci-Tech Daresbury, Keckwick Lane, Daresbury, Warrington WA4 4AD, United Kingdom}
\affiliation{Department of Physics, Lancaster University, Bailrigg, Lancaster LA1 4YB, United Kingdom}

\author{Darren M. Graham}
\affiliation{The Cockcroft Institute, Sci-Tech Daresbury, Keckwick Lane, Daresbury, Warrington WA4 4AD, United Kingdom}
\affiliation{Department of Physics and Astronomy \& Photon Science Institute, The University of Manchester, Oxford Road, Manchester M13 9PL, United Kingdom}

\author{Morgan T. Hibberd}
\affiliation{The Cockcroft Institute, Sci-Tech Daresbury, Keckwick Lane, Daresbury, Warrington WA4 4AD, United Kingdom}
\affiliation{Department of Physics and Astronomy \& Photon Science Institute, The University of Manchester, Oxford Road, Manchester M13 9PL, United Kingdom}

\author{Steven P. Jamison}
\affiliation{The Cockcroft Institute, Sci-Tech Daresbury, Keckwick Lane, Daresbury, Warrington WA4 4AD, United Kingdom}
\affiliation{Department of Physics, Lancaster University, Bailrigg, Lancaster LA1 4YB, United Kingdom}

\author{Robert B. Appleby}
\affiliation{The Cockcroft Institute, Sci-Tech Daresbury, Keckwick Lane, Daresbury, Warrington WA4 4AD, United Kingdom}
\affiliation{Department of Physics and Astronomy \& Photon Science Institute, The University of Manchester, Oxford Road, Manchester M13 9PL, United Kingdom}

\date{\today}% It is always \today, today,

\begin{abstract}
The short wavelength of terahertz (THz) waves makes the acceleration of electrons, from typical electron gun energies to fully relativistic energies, challenging due to the need to match the velocities over comparably long distances. However, the inherent laser synchronisation of a photogun to the THz drive laser offers significant advantage in applications where timing is critical. We present a novel design process for high-gradient THz-driven dielectric-lined waveguide injectors and demonstrate designs for rectangular and cylindrical geometries. These structures utilize novel tapering and stepping schemes to manipulate the phase velocity, controlling the position of the electron bunch relative to the phase of the accelerating field. We demonstrate a hybrid optimisation method utilizing firstly a multi-objective genetic algorithm based on analytic models of the waveguide accelerating modes, and then high-detail particle-in-cell simulations. Our example designs achieve electron acceleration from 100\,keV to 1\,MeV over a distance of 20\,mm by interacting with a multicycle 0.5\,mJ, 0.2\,THz pulse, paving the way for the realisation of THz-based injectors. The exit beam shows excellent beam quality with 50\,fs bunch lengths, sub 0.5\% energy spread and emittances of under 0.15\,\textmu rad. We also present an analysis of the robustness of the designs to errors in machining and operational parameters demonstrating the feasibility of the concept. We show that using multi-objective genetic algorithm optimization and robust design process, we can achieve high-quality 1\,MeV beams.
\end{abstract}

\maketitle

%\tableofcontents

\section{\label{sec:Intro}Introduction }

The operational frequency of conventional accelerators imposes significant limitations on achievable bunch phase space modulation. A number of schemes utilizing higher-frequency electromagnetic fields have been investigated. High gradients (beyond 500\,$\mathrm{MeV\,m^{-1}}$) have been demonstrated by laser-induced acceleration in dielectric micro-structures both in the relativistic and subrelativistic regimes \cite{DLA, DLA2, DLA3}. Although optical frequencies enable higher gradients, they also correspond to extremely short sub-micrometre wavelength tolerances in the manufacturing of accelerating structures and to a limited beam acceptance radius. Operating at low THz frequencies (mm-scale wavelength) offers a good middle ground between fabrication tolerance using nano computer numerical control (CNC) machining \cite{CNC} and accelerating gradients, while still offering picosecond-scale control over bunch properties \cite{THz_electron_manipulation, STEAM}.

Typical THz-driven acceleration techniques are based on the interaction between an electron bunch and a travelling THz pulse; hence, a high-power THz source is essential for operation. At first, experiments relied on high-current electron beams inducing THz-frequency wakefields in dielectric-lined structures reaching accelerating fields exceeding 1\,$\mathrm{GeV\,m^{-1}}$ \cite{wakefield_DLW_OShea2016} at single-shot operation showcasing potential for reaching higher gradients. The last few decades have brought major advances in THz-generation technology, especially laser-driven sources, such as optical rectification in lithium niobate (LiNbO$_3$) \cite{Oh2014, Cliffe2016, Fulop2020}. Methods have been developed not only to increase THz electric field amplitudes but also to reduce spectral bandwidths, extending pulse lengths and increasing the interaction length, which is limited by significantly slower group velocity compared to phase velocity in the considered THz structures. A method of generating these multi-cycle pulses is, instead of a single periodically-poled lithium niobate (PPLN) crystal, to use stacks of PPLN wafers \cite{lemery2020PPLN, Mosley2023}. This allows generation of narrow-band pulses of hundreds of microjoules \cite{ELI_PPLN_highE} that excite accelerating modes over extended interaction lengths, e.g. in dielectric-lined waveguides \cite{cylindDLWtaper, Hibberd2020, Nix2024, Guarnera2025} or mm-scale cavities \cite{THz_cavities}. 

Laser-driven THz acceleration has been demonstrated in a cylindrical dielectric-lined waveguide (DLW) \cite{Nanni2015}, where an optically generated THz pulse excited the cylindrically symmetric TM$_{01}$ accelerating mode. THz-driven acceleration was also demonstrated in a rectangular DLW structure \cite{Hibberd2020}, which supports a class of hybrid modes known as longitudinal-section (LS) modes. For acceleration, the longitudinal-section magnetic mode, LSM$_{11}$, provides a strong longitudinal electric field component \cite{DLWmodes1, DLWmodes2}.

In the relativistic acceleration regime, particle velocity remains roughly equal to the speed of light throughout, so selecting a geometry with phase velocity $v_p \approx c$ will accelerate with little to no phase slippage. In the subrelativistic regime, however, the particle velocity increases significantly throughout the interaction and would therefore quickly slip out of phase alignment in a constant-phase-velocity structure. To maintain acceleration over prolonged distances, it is necessary to correct for this slippage. Several schemes have been proposed to address this issue, including the use of phase shifters \cite{phase_shifters}, multi-stage acceleration \cite{multi_stage_DLW}, and tapered geometry \cite{Nix2024, cylindDLWtaper}. By tailoring the cross-sectional geometry of the DLW along its longitudinal direction, the phase velocity of the accelerating waveform can be adjusted to maintain synchronism with the accelerated bunch. The geometry that provides the perfect synchronization between the THz waveform and a reference particle takes a complex, slowly varying adiabatic form \cite{cylindDLWtaper}, which is challenging to manufacture. However, finite bunch size and the focusing properties of the accelerating mode at various phases with respect to the accelerating crest mean that perfect synchronization is not necessary, and controlling the phase-alignment through velocity mismatches is a viable path towards stable acceleration similar in principle to alternating phase focusing linac \cite{Good1953PhaseReversal, Fainberg1956}.

In our previous study \cite{Nix2024, NixLINAC}, we demonstrated a design framework for tapered rectangular DLWs driven by THz pulses achievable using a typical commercial ultrafast amplifier accelerating sub-relativistic beams. In this paper, we show a route to 1 MeV over a single novelly tapered structure with a gradient over 20 times higher compared to the previous study. The described technique enables exceptional control of both the longitudinal and transverse phase space of sub-picosecond electron bunches, making it attractive for ultrafast science applications \cite{THzbasedUED}, and as a compact booster stage for higher-energy linacs operating in the relativistic regime.

Here, we take a starting energy of 100\,keV ($v=0.548 \,c$), easily accessible with commercial DC photoelectron guns. Parts of the core design process remain applicable, but new difficulties arise in this higher gradient regime. To cover the large velocity range in a single tapered structure poses a challenge: if changing width alone as in \cite{Nix2024}, the DLW cannot be optimal at both the weakly relativistic and highly relativistic sections. In this paper, we demonstrate how this problem can be overcome using alternative tapering schemes that include changes in dielectric thickness. We discuss the electromagnetic properties of DLWs to compare the potential performance of thickness-tapered and width-tapered rectangular DLWs. We also consider radially width-tapered cylindrical DLWs. We introduce a design process involving optimisation via a genetic algorithm that can be applied to any set of operating parameters, tolerance limits, or relative weightings of design goals, such as energy, transverse beam emittance, and energy spread. We present demonstrative designs in both rectangular and cylindrical geometries and discuss the achieved performance and the required tolerances for machining and excitation-pulse parameters.

\section{Waveguide geometry \& beam dynamics}

This section discusses the impact of the cross-sectional geometry of DLWs on their supported accelerating modes, and the consequences this implies for the dynamics of the accelerated beam depending on its longitudinal phase with respect to the accelerating waveform.

\subsection{Rectangular waveguide}
\begin{figure}[hbt!]
    \centering
    \includegraphics[width=8.6 cm]{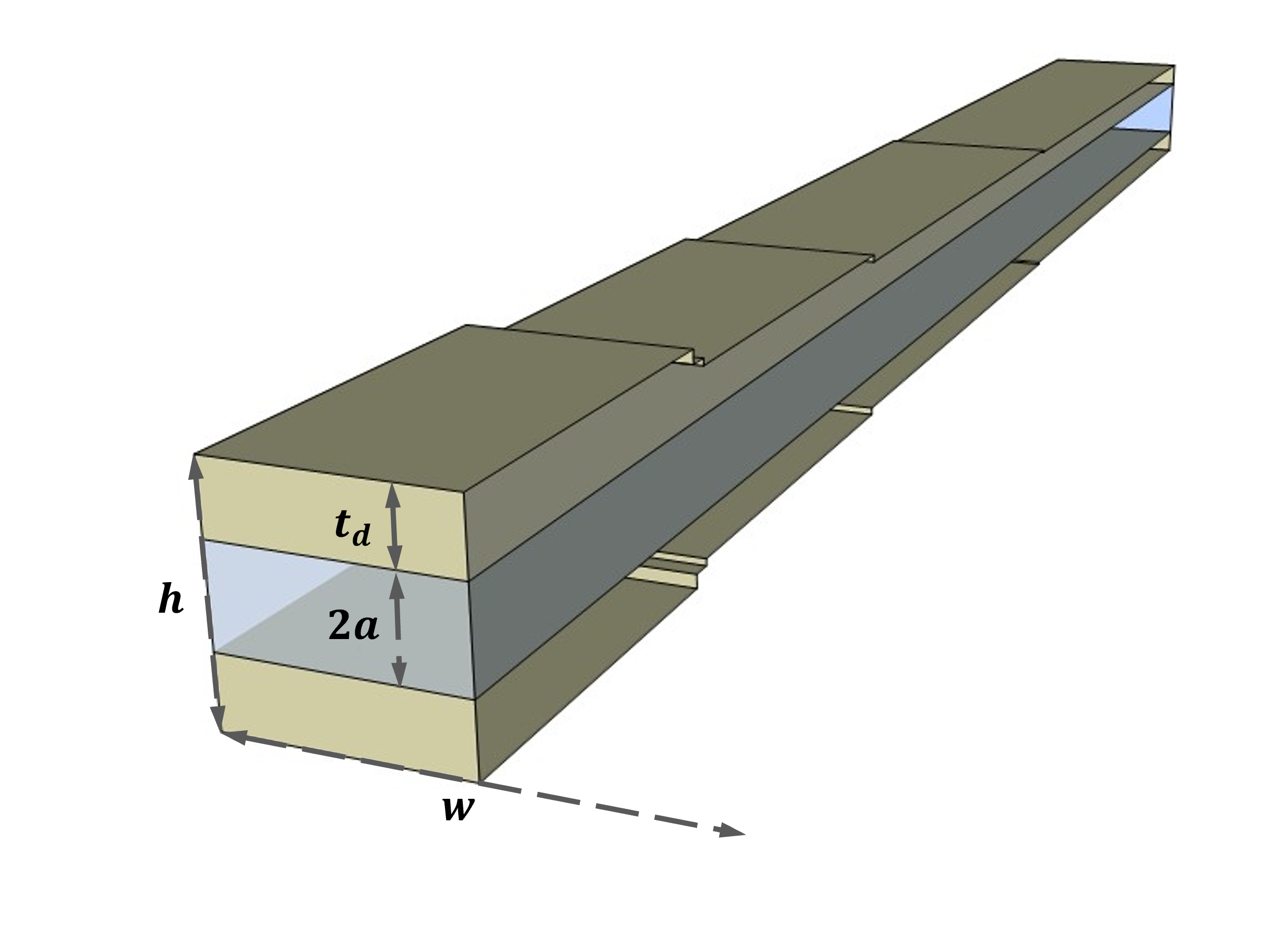}
    \caption{Section of rectangular dielectric-lined waveguide (of width $w$, height $h$, and vacuum gap $2a$) lined with dielectric lining of thickness $t_d$ embedded in conductive channel.}
    \label{fig:dlw_rect}
\end{figure}

The rectangular DLW consists of a metal waveguide with width $w$ and height $h$. Its top and bottom walls are lined with a dielectric layer of thickness $t_d$, width $w_d$ and dielectric constant $\epsilon_r$. The mode of interest for acceleration is the longitudinal section magnetic mode, $LSM_{11}$. The dispersion relation states that the phase velocity at which this mode propagates can be varied by adjustment of DLW parameters, illustrated in FIG. \ref{fig:dlw_rect}. As the height of the structure has little impact on the phase velocity $v_p$ for thick dielectrics \cite{Nix2024}, and we constrain our choice of dielectric to an isotropic and homogeneous material, we consider only adjusting the width of the waveguide and the thickness of the dielectric lining (keeping vacuum gap height $2a$ constant, so also varying total height $h$). For both $w$ and $t_d$, the decrement in value leads to higher phase velocity. Tapering either or both of these parameters can therefore provide the required velocity matching for extended interaction.

The field expressions of $LSM_{11}$ in non-tapered rectangular DLWs have previously been derived analytically using the mode-matching method along with the dispersion relation \cite{analyticDLW} that can be solved numerically. Here, we propose that under the assumption that the local geometry determines the EM field distribution (valid for slowly-tapered waveguides) and only $LSM_{11}$ is excited, the field expressions can be modified by the introduction of $z$-dependency derived from longitudinal variation of width and thickness: $w(z)$ and $t_d(z)$. The dispersion relation can be expressed as:

\begin{equation}
     \epsilon_r k_{y}^{0} \cot(k_{y}^{0}a)
    - k_{y}^{1} \tan \left(k_{y}^{1}(t_d(z))\right) = 0,
    \label{eq:dispersion}
\end{equation}

where the vacuum wavenumber $k_0 = \omega/c$ determines

\begin{equation}
     k_{y}^{0} = \sqrt{k_0^2 -k_z^{2}-k_x^2},
\end{equation}
 \quad and
\begin{equation}
     k_{y}^{1} = \sqrt{\epsilon_rk_0^2 - k_z^{2} - k_x^2},
\end{equation}
\quad and
\begin{equation}
     k_{x}=\frac{\pi}{w(z)}.
\end{equation}

Equation (\ref{eq:dispersion}) can be solved for longitudinal wavenumber $k_z(z)$ determining the phase velocity in longitudinal direction, $v_p(z) = \omega /k_z(z)$.

As the phase velocity of the propagating mode increases down the structure, the waveform is effectively longitudinally stretched as it propagates, which can be represented by replacing the longitudinal field dependence term of the non-tapered waveguide with an integral term:

\begin{equation}
     exp\left(-i k_z z \right) \rightarrow \exp\left(-i \int_{0}^{z} k_z(z) \, dz \right).
\end{equation}

Additionally, as the width and the dielectric are tapered, the cross-section of the guide decreases, increasing the energy density and corresponding to higher field amplitudes. Therefore, amplitude also becomes $z$-dependent. The exact amplitude at a given $z$ can be found by considering the conservation of energy flux (power flow), given by the time-averaged Poynting vector integrated over the cross-sectional area in both the dielectric and vacuum regions. Therefore:

\begin{equation}
     P = \frac{1}{2}\iint \vec{E} \times \vec{H}^*  \, \cdot \hat{z} \ dx \ dy = -\frac{1}{2}\iint E_y H_x^* \ dx \ dy .
\end{equation}

Assuming a lossless waveguide, the above relationship can be directly used to compute the amplitude factor $A(z)$  at any point along the waveguide by normalizing it with respect to constant power flowing through its cross-section:

\begin{equation}
      P\propto A^2(z)  .
\end{equation}

Modifying the established $LSM_{11}$ mode formula by this longitudinal dependence gives the analytic field equations in vacuum:

\begin{equation} \label{eq:fields_rect}
\begin{aligned}
E_x &= -A(z) k_x k_{y}^0 \sin \left( k_x x \right) \cos \left(k_{y}^0 y \right) ,\\
E_y &= A(z) ( k_x^2+ k_z^2) \cos \left( k_x x \right) \sin \left(k_{y}^0 y \right) ,\\
E_z &= A(z)k_z(-ik_{y}^0) \cos \left( k_x x \right) \cos \left(k_{y}^0 y \right) ,\\
H_x &=  -A(z) \omega \varepsilon_0 k_z \cos \left( k_x x \right) \sin \left(k_{y}^0 y \right)  ,\\
H_y &= 0 ,\\
H_z &=  -A(z)i\omega \varepsilon_0 k_x \sin \left( k_x x \right) \sin \left(k_{y}^0 y\right) ,\\
\end{aligned}
\end{equation}

\noindent where each term is multiplied by a travelling wave expression with a phase offset $\phi$:

\begin{equation}
    \exp\left(i\omega t - i \int_{0}^{z} k_z(z) \, dz  + i\phi \right).
\end{equation}

The above field equations have been verified against the equivalent CST Studio Suite numerical simulation of a gradually tapered rectangular DLW. %\ref{fig:dlw_analytic_cst})

\begin{figure}[h]
    \centering
    \includegraphics[width=8.4 cm]{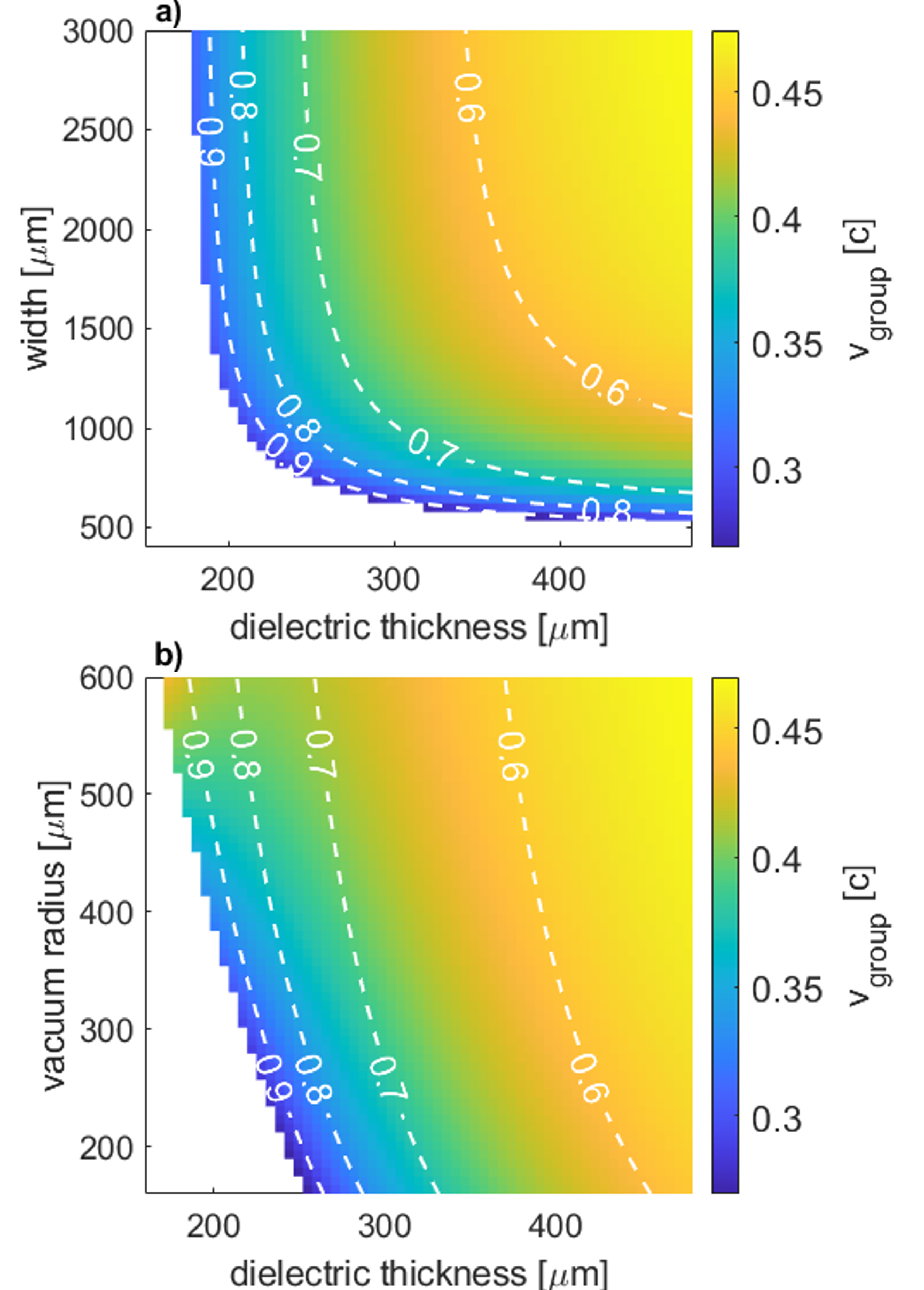}
    \caption{\label{fig:vp_vg_comparison}{Rectangular $LSM_{11}$ at $a = 200$\,\textmu m (a) and cylindrical $TM_{01}$ (b) phase (white contour) and group (color) velocities for various geometries at $f = 0.2$ THz. The white region corresponds to phase velocities greater than $c$.}}
\end{figure}

The strong longitudinal component of mode $LSM_{11}$, supported by the rectangular waveguide, enables acceleration. However, its transverse profile is not flat. As the dielectric lining increases the longitudinal wavenumber, slowing the propagation of the mode in the structure, the non-zero transverse wavenumbers result in a strong non-linear transverse dependence of $E_z(x,y)$. This manifests primarily as a quadrupole component \cite{orthoDLW}, in which the vacuum $E_z(x,y)$ rises, reaching a maximum at the dielectric lining boundary $y \rightarrow \pm a/2$, and drops to zero at conductor walls $x \rightarrow \pm w/2$ leading to a transversely correlated energy spread of the bunch accelerated at constant phase. Similarly to classic accelerating structures, a bunch ahead of the accelerating crest, $\phi\in(\pi,\frac{3}{2}\pi)$, experiences longitudinal focusing, as the tail of the bunch experiences a higher field than its head. Respectively, the bunch positioned behind the crest, $\phi\in(\frac{1}{2}\pi,\pi)$, experiences longitudinal defocusing.

The variance of the $LSM_{11}$ electric longitudinal field across the transverse plane gives rise to transverse fields as a result of the Panofsky-Wenzel theorem \cite{PanofskyWenzel}. The dominating components of that fields are monopole and quadrupole. Hence, the transverse field is 0 at the central axis and $|E_x|$, and $|E_y|$ increase with $x \rightarrow \pm w/2$, and $y \rightarrow \pm a$ respectively. These amplitudes of transverse electric fields are strongly related to the waveguide width through $k_x$, and the wider waveguide results in their lower strengths. Additionally, lower phase velocities are determined by higher $k_z$, disproportionately strengthening the electric field in the $ y$-axis compared to the $ x$-axis. This has a profound impact on the beam as particles experience strong focusing in the $y$-axis and weak defocusing in the $x$-axis when behind the accelerating field crest, $\phi\in(\frac{1}{2}\pi,\pi)$, and accordingly strong defocusing in the $y$-axis and weak $x$-focusing when ahead of it,  $\phi\in(\pi,\frac{3}{2}\pi)$. Therefore, the main objective of transverse optimisation is to mitigate the blow-up in the $y$-dimension. This can be achieved by staying at a constant phase behind the bunch; however, it is not an optimal solution due to the longitudinal bunch defocusing that occurs at this phase.

As the longitudinal field experienced by the bunch is strongly dependent on the transverse coordinates, and the transverse forces acting on the particle are dependent on its longitudinal position with respect to the accelerating crest of the waveform, the beam dynamics exhibit a strong longitudinal-transverse coupling. The coupling is especially evident between transversely dominating $E_y$ and longitudinal $E_z$ fields, as $y$-directed deflection results in a higher longitudinal field $E_z(y)$ leading to higher phase advance and even stronger deflection. To mitigate the effects of this coupling, an acceleration scheme must maintain balance by keeping the bunch in the proximity of the longitudinal axis, while mitigating non-linearities in transverse fields and, hence, emittance blow-up. This allows alternating defocusing and focusing forces to act in the $x$-, $y$-, and $z$-directions while accelerating the bunch.

\subsection{Cylindrical waveguide}

Another well-established geometry choice is the cylindrical DLW, where the conducting guide, of radius $R$, is lined with a dielectric of thickness $t_d$ and constant $\epsilon_r$, leaving a vacuum aperture of radius $r_0=R-t_d$. The fundamental mode with a strong longitudinal component is $TM_{01}$. The dispersion relation states that the phase velocity at which this mode propagates can be varied by adjustment of DLW parameters: $r_0$, $t_d$, $\epsilon_r$. However, FIG. \ref{fig:vp_vg_comparison} shows that, similar to the rectangular configuration, changes in dielectric thickness have a much greater effect on phase velocity than variations in the vacuum gap radius. Hence, when optimising the cylindrical DLW, we fixed the diameter at the same value as the rectangular waveguide vacuum height and varied only the dielectric thickness. This mode was previously analysed, and an adiabatically tapered structure was designed to provide synchronous acceleration \cite{cylindDLWtaper}. The fields in such a mode take the following form in vacuum ($r<r_0$):

\begin{figure}[hbt!]
    \centering
    \includegraphics[width=8.6cm]{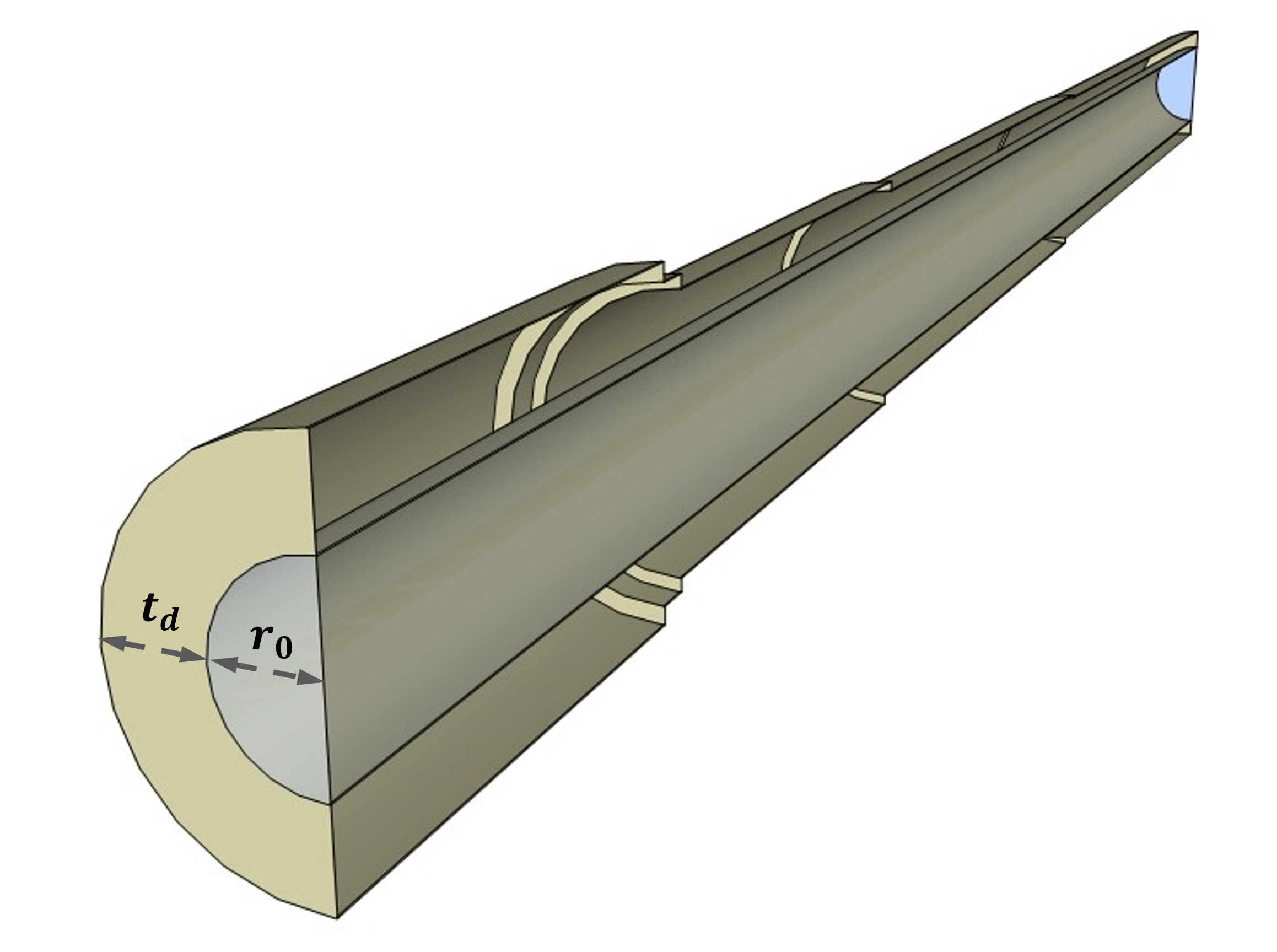}
    \caption{Section of cylindrical dielectric-lined waveguide (with vacuum gap of radius $r_0$ and dielectric lining of thickness $t_d$) embedded in a conductive channel.}
    \label{fig:dlwcylind}
\end{figure}

\begin{equation}
\begin{aligned}
E_{z} &= A_z(z) I_{0}\big(r k_{r}(z)\big)  ,\\
E_{r} &= \frac{A_z(z) k_{z}(z)}{k_{r}(z)} I_{1}\big(r k_{r}(z)\big)  ,\\
B_{\phi} &= \frac{A_z(z)\omega \epsilon_{0} \mu_{0}}{k_{r}(z)} I_{1}\big(rk_{r}(z)\big) ,\\
\end{aligned}
\end{equation}

\noindent where the radial propagation constant is expressed as $k_{r} = \sqrt{k_0^2 - k_z^{2}}$, and each term is multiplied by a travelling wave expression with phase offset $\phi$:

\begin{equation}
    \exp\left(i\omega t - i \int_{0}^{z} k_z(z) \, dz  + i\phi \right).
\end{equation}

The radial dependence of the field is determined by modified Bessel functions of the first kind of order zero and one, consecutively $I_{0}$ and $I_{1}$. 

The $TM_{01}$ also exhibits a strong longitudinal component. Like $LSM_{11}$, the electric field magnitude increases closer to the dielectric lining. The radial dependence of the longitudinal field is closely related to the transverse fields defocusing the bunch. Ahead of the accelerating crest, $\phi\in(\pi,\frac{3}{2}\pi)$, it experiences longitudinal focusing as the tail of the bunch experiences a higher field than its head. Within the same phase region, the transverse fields act on the bunch, resulting in transverse defocusing. Respectively, the bunch positioned behind the crest, $\phi\in(\frac{1}{2}\pi,\pi)$, experiences longitudinal defocusing and transverse focusing accordingly. Hence, in the cylindrical case, the phase choice always favours stability in either longitudinal or transverse direction.

\section{Geometry optimisation process \label{sec:geom_opt}}
As shown in the previous section, the phase velocity can be varied by decreasing both the width and the dielectric lining thickness of the waveguide. This section presents a set of considerations for finding the geometry of a structure capable of accelerating a 100 keV electron bunch to 1 MeV energy ($v = 0.94 \,c$) while conserving bunch quality. The focus of the discussion is the relationship between cross-sectional geometry and EM fields determining the beam dynamics.

\subsection{DLW excitation \& frequency choice}

\begin{figure}[h]
    \centering
    \includegraphics[width=8.6 cm]{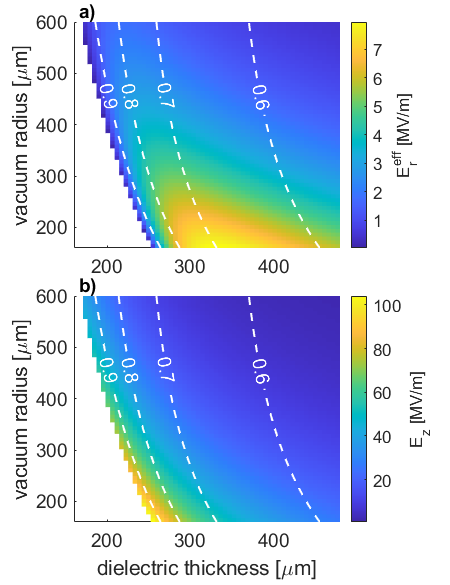}
    \caption{\label{fig:efields_c}Cylindrical DLW electric field amplitudes. a) radial $r = a/2 = 100$\,\textmu m off-axis; and c): longitudinal on-axis. The contours correspond to phase velocity [c] at a given geometry. The radial field amplitude is corrected by the opposing impact of the magnetic field: $E_r^{eff}=E_r-v_p B_{\phi}$. }
\end{figure}

As the group velocity of $LSM_{11}$ and $TM_{01}$ modes at subluminal phase velocity is significantly lower than the velocity of the 100 keV bunch, the accelerating DLW must be excited by a narrowband (multi-cycle) source to deliver a substantial interaction length within the structure. For the design simulations here, we consider a long, flat-topped pulse that can be feasibly generated in laser-driven PPLN sources \cite{Jolly2019}. The choice of source frequency determines not only the waveguide spatial dimensions but also the maximum achievable accelerating gradient for a THz source of given power. Higher frequency offers potential to elevate the breakdown limit of the structure and support greater accelerating fields \cite{DLW_breakdown}. Additionally, under the assumption that the dielectric permittivity does not change substantially over a wide sub THz frequency range, the dispersion relation becomes invariant under simultaneous scaling of the structural dimension and inversely proportional scaling of the frequency. Scaling structure this way means that the field amplitude is linearly proportional to frequency at a given power fed to the structure. For example, in a rectangular DLW, we can double the on-axis electric field by doubling the frequency and halving the DLW dimensions ($a$, $w$, $d$), while keeping their ratio $a:w:d$ constant. However, this comes at the cost of shorter wavelengths, and hence greater field variation across the length and diameter of the electron bunch. Moreover, simply halving $a$ is not always feasible, as it must be chosen to allow suitable beam clearance; under this constraint, higher frequency no longer necessarily corresponds to higher gradient. For our illustrative designs, 0.2\,THz was chosen as the operating frequency, which corresponds to a 1.5\,mm wavelength, considerably longer than the micrometre-scale bunches considered in this work. The vacuum gap was kept constant at $2a = 400$\,\textmu m to balance high accelerating fields with transverse deflecting fields and to provide clearance for the resulting bunch expansion. For the same reasons clearly visible in FIG \ref{fig:efields_c}, the vacuum radius of cylindrical waveguide was set at the same value $r_0=a=200$\,\textmu m. Although a single frequency is considered, geometrical relationships and design principles are applicable for a wide range of frequencies. In this work, we consider excitation with up to 0.54\,mJ of energy within a narrowband 150\,ps pulse centred at 0.2\,THz. We optimize for a constant time-averaged power of 3.6\,MW, and the actual minimum required pulse duration (and hence pulse energy) depends on the integrated group velocity of the structure. This energy was set as an upper limit; in practice CST simulations suggest that the cylindrical waveguide can be fully excited with shorter pulse of 135\,ps corresponding to 0.49\,mJ energy.

\subsection{Field distribution}
\begin{figure}[h]
    \centering
    \includegraphics[width=8.6 cm]{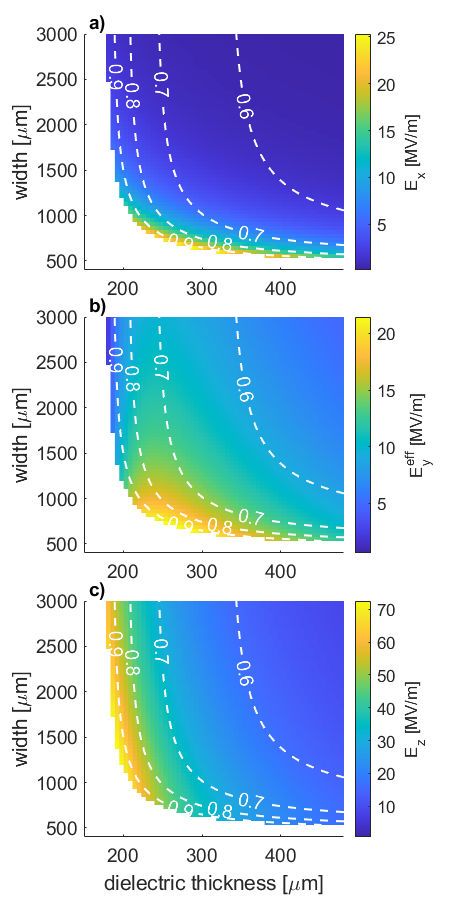}
    \caption{\label{fig:efields}Rectangular DLW electric field amplitudes in . a) and b): transverse $(x,y) = (a/2, a/2)$ off-axis; and c): longitudinal on-axis. The contours correspond to phase velocity [c] at a given geometry. The amplitude in the y-direction is corrected by the opposing impact of the magnetic field: $E_y^{eff}=E_y-v_p B_x$. }
\end{figure}

The boundary conditions in the waveguide govern the spatial field distribution. The cross-sectional geometry determines the mode wavenumbers in transverse and longitudinal directions. These directly influence the EM field magnitudes through the equation (\ref{eq:fields_rect}). As shown in FIG. \ref{fig:vp_vg_comparison}, both $w$ and $t_d$ can be manipulated to increase the phase velocity. However, varying each has a distinct impact on the wavenumbers and EM field distribution in the vacuum. To investigate this impact, the electric field magnitude in vacuum was computed for a wide range of configurations of specific width and dielectric thickness FIG. \ref{fig:efields}. This was done by normalizing it to the energy flux flowing through the cross-section. The figure shows that narrower waveguides are clearly related with higher transverse fields. Moreover, the x-oriented magnetic field impact on particles naturally opposes y-oriented electric field. This cancellation is especially visible when luminal velocities are approached in wider waveguides shown in FIG. \ref{fig:efields} b). Thicker dielectric layers confine the modal energy to the lining, resulting in lower on-axis longitudinal fields; however, they are also necessary to achieve lower phase velocity. This was addressed in our previous work in the subrelativistic regime \cite{Nix2024} by injecting ahead of the crest in a velocity-mismatched region, allowing for a thinner dielectric at the beginning than would be required to exactly match the initial particle velocity, and we employ the same principles here.

A taper in the structure can be represented as a line (or a curve) between two points in FIG. \ref{fig:vp_vg_comparison}. A vertical line is a tapering of the width, and a horizontal line is a tapering of the thickness. In either case, the sensitivity to the tapered variable increases as the variable gets smaller, as seen by the closeness of the contour lines. Our previous study \cite{Nix2024} demonstrated the effectiveness of width-tapering in the subrelativistic regime, using a constant dielectric thickness for a relatively low-energy pulse. Here, we consider much larger pulse energies (and hence higher gradients), thus covering a large energy range in one stage. Applying the same method of width-tapering alone would depend on machining parts with extreme tolerance. A thinner dielectric lining could alleviate this, but would have a higher minimum phase velocity, unsuitable for the low velocity end of the device. Therefore, it can be concluded that to achieve good performance at both the weakly relativistic velocities and highly relativistic velocities, the dielectric width must change somewhere along the interaction length. In \cite{Nix2024}, we discuss this problem and the possibility of solving it using multistage acceleration, in which a series of tapered structures, each with a thinner dielectric layer than the previous, is used to reach high energy. That approach would be effective for low-gradient subrelativistic stages, but here, we consider a higher-gradient regime, in which there is sufficient pulse energy to reach 1\,MeV in a single structure. To effectively utilize that much pulse energy in one stage, the thickness change must happen within the single structure. This does also come with a manufacturing challenge, as sensitivity to thickness is also high in the high-velocity region, and any jump in thickness risks alignment issues and reflection, but these are not insurmountable barriers to the design. Section III details the field distribution for different combinations of width and thickness, which can be used to determine which tapering scheme to use, and we consider the level of reflection that occurs over steps in thickness.

\subsection{Adiabatic, Tapered, and Stepped Solutions}

In FIG. \ref{fig:efields}, it is shown that for a specific phase velocity, the highest accelerating field tends to be found in small-thickness, large-width waveguides. The best solution, therefore, is a thickness-tapered DLW. This option was rejected in our previous work due to a desire for simpler machining by cutting a taper from a single sheet. We return to the manufacturability question later in this section, but first temporarily set it aside for the purpose of discussing the implied ideal solution of an adiabatically tapered thickness. Given the constant vacuum height, frequency and dielectric constant ($2a$, $f$, $\epsilon_r$), the width $w(z)$ and the dielectric thickness $t_d(z)$ can be adiabatically tapered to provide phase synchronization for a synchronous particle throughout the structure. The taper can be found by setting the initial amplitude of the field $A(z=0)$ (normalized with respect to the modal power) and final synchronous particle velocity (energy), then integrating the longitudinal equation of motions while solving the dispersion relation at each step to provide particle-wave synchronism, similarly to \cite{cylindDLWtaper}. Although the synchronous adiabatic solution minimizes power reflections within the structure, it has two major drawbacks. The first one is the fact that synchronization with phase is provided only for a single synchronous particle injected on-axis, right at the crest of the accelerating field $(x_0,y_0,\phi_0) = (0,0,\pi)$. This means the particles occupying space in a real 3D bunch will experience varying accelerating fields due to the longitudinal and transverse dependence of the accelerating field $E_z(x,y,\phi)$. This can be mitigated by driving the tapered structure with a higher field than it was designed for and bunch injection ahead of the accelerating crest, providing longitudinal stability, but setting the bunch in a transversely defocussing phase. The second issue arises from the fact that the adiabatic taper takes a complex shape, thus requiring very high manufacturing precision. Especially as the phase velocity approaches the speed of light, the phase velocity becomes very sensitive to even sub-micron-level dielectric thickness changes. Therefore, we propose an approximation of a dielectric thickness taper by discretizing it with N segments of constant thickness, so that for $N \rightarrow \infty$ it converges to the adiabatic form. Many smaller steps tend to result in less reflection loss than a few large steps, but each step is an additional interface where coherent reflections and manufacturing imperfections, such as misalignments and small gaps between segments, can occur. Therefore, we elected to use as few sections as we could without reflection loss or phase slippage becoming a major issue. Since we propose a series of segments of constant phase velocity, the electrons will not stay perfectly synchronized to the crest, but rather drift back and forth around the crest. This alternating phase-focused approach can be beneficial, as it means the bunch spends some of its time in a transversely focusing phase and some of its time in a longitudinally focusing phase. Further precise control may be achieved by tapering the widths within each section, but in this case, the straight-sectioned step model provided very good performance, as shall be shown in the following sections, and this extra refinement was not considered worthwhile as it would provide very marginal benefits at the cost of significantly increased complexity of optimisation and manufacture.

\begin{figure}[hbt!]
    \centering
    \includegraphics[width=8.6 cm]{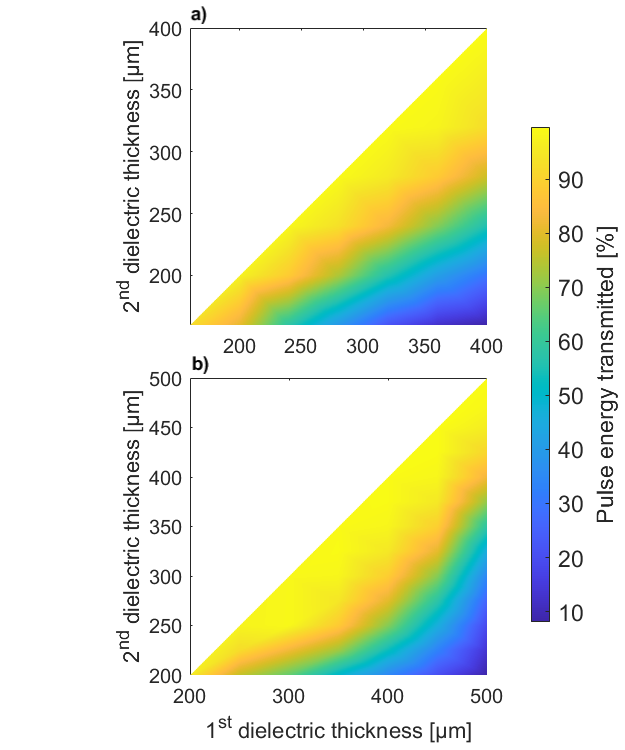}
    \caption{The energy transmission across a step in dielectric thickness for a): rectangular waveguide, and b): cylindrical waveguide simulated in CST Studio Suite. In both cases, the gap height was kept constant, the outer dimension stepped, and an 8-cycle 0.2\,THz signal was used for the excitation.}
    \label{fig:stepreflect}
\end{figure}

In the adiabatic case, reflection losses as the THz pulse propagates along the taper are very low due to the lack of any abrupt transitions. Changing to steps between different thicknesses necessitates that the model considers reflections at each interface. A simulation was carried out examining a finite pulse propagating across the geometry step, repeated for many combinations of step size, as illustrated in FIG. \ref{fig:stepreflect}. By taking the expected energy reflection, the impact of the reflection on accelerating fields can be estimated. This quantity, however, is a slight approximation, as the numbers in FIG. \ref{fig:stepreflect} all refer to the specific case of a short, 8-cycle pulse, while the exact reflection value will vary slightly for different pulse bandwidths. Despite the approximation, these figures were observed to provide sufficient agreement between the fast optimizer model and the particle-in-cell simulations.

FIG. \ref{fig:stepreflect} considers only the case where the vacuum gap height is constant, and the step is in the outer height. The figure demonstrates that, if step sizes remain fairly small, nearly all of the energy is transmitted. An alternative geometry was considered in which the total waveguide height was fixed, and therefore, the steps increased the gap height along the length. In this regime, the reflections were significantly lower, but as the on-axis field decreases with increasing gap height, the resulting field after a step was typically still lower than the constant-gap-height equivalent in the ranges relevant to our design. This ceased to hold for small steps between low-thickness dielectrics; however, for the purposes of our illustrative design, we neglected this added complexity because the potential benefit was not deemed worth the increase in the working parameter space for optimisation. \par

Steps in the dielectric thickness were observed to induce a small phase shift in the forward signal due to both the capacitance of the step and the superposition of reflected waves over longer timescales, both proportional to the step height. The capacitance of the step gives both a transmission phase shift and a reflected wave. The phase shift in this case varies roughly linearly with the step height and is around $8\degree$ when the step height is 40\% of the first dielectric thickness. From this, we found that replacing a single step two smaller steps reduced the total phase shift. As the later steps in our design are smaller and therefore cause lower reflection, this was applied only between the first and second constant dielectric DLW segments. For our design, this lowered a $7\degree$ shift to around $4\degree$.
An additional time-dependent phase shift effect can also occur in long-pulse regimes due to interference between the primary forward signal and the component that is reflected backwards by a step, then forwards again by an earlier step with the phase shift proportional to the distance between the steps. This was found to shift the phase by up to $\pm2\degree$ relative to the single-step shifts, a small enough range to neglect from the design optimisation. 
In rectangular case specifically, sudden step discontinuity may adversely affect the accelerating mode not only through reflection, but also through conversion to other modes (primarily LSE$_{11}$, which has a longitudinal field distribution similar to LSM$_{11}$). This coupling has been found to be weak in the considered geometries, and combined with the fact that LSE$_{11}$ travels at much higher phase velocities than LSM$_{11}$, it has not significantly interfered with the dominating accelerating mode. These discontinuity effects alter performance on a small scale, similar to the other approximations used, and hence it is reasonable to neglect them during the initial optimisation routine for simplification; sufficient agreement between the analytic and particle-in-cell models was found without considering them during optimisation.
\par

In summary, FIG. \ref{fig:efields} \& \ref{fig:stepreflect} confirm that there exists a relatively wide tolerance gap in terms of reflected signal, and that wide and thin-lining DLWs tend to have improved field parameters over narrow and thick-lining ones at the same synchronous velocity, which opens up a broader design space. In our previous work \cite{Nix2024}, we elected to only taper the width for manufacturing reasons, limited by what was readily available for immediate machining. Now we consider the wider design space with steps in thickness as the primary means of controlling phase velocity. We apply the same principle to the cylindrical geometry option. Hence, we define the structure taper problem in a space of finite dimension defined by number of steps.

\subsection{\label{sec:GA}Optimisation using Genetic Algorithm}
As the problem space becomes high-dimensional, finding desirable solutions becomes less trivial, but certain algorithms have been shown to perform well in multi-dimensional spaces. In this work, we employ a multi-objective genetic algorithm (MOGA) to generate structures optimized for user-defined goals. A genetic algorithm is an optimisation method inspired by natural selection, where a population of candidate solutions evolves over generations through selection, crossover, and mutation to find near-optimal solutions. Each candidate is evaluated according to a set of objective functions, so better-performing individuals are more likely to contribute to subsequent generations. In a multi-objective genetic algorithm, competing design objectives can be considered simultaneously without combining them into a single weighted cost function. Instead, the algorithm identifies a set of non-dominated solutions forming a Pareto front, allowing the direct assessment of trade-offs between different design goals.

The MOGA (NSGA-II MATLAB default \cite{GA_NSGA-II}) has been implemented by wrapping it around a custom field particle tracker (developed in MATLAB) based on the 4\textsuperscript{th} order Runge-Kutta method and analytic fields derived in each segment. The number of optimisation objectives was constructed based on accelerated bunch qualities such as final charge capture rate CR (determining the fraction of particles that have not been deflected and hence lost), and its projected RMS spread $\Delta E_{RMS}$, as well as transverse RMS emittance $\epsilon_{xy}$ (defined as geometrical mean of projected normalized $x$- and $y$- emittances). To ensure each solution delivers a bunch with desired energy of at least 1\,MeV, the non-linear penalty was applied whenever the bunch fell out of the $1.05\pm0.05$\,MeV. The multi-objective optimisation approach allowed us to find sets of solutions converging to Pareto-optimal. Each characteristic of the Pareto-superior solution cannot be improved without compromising others. For both the cylindrical and rectangular design options, a single example of a well-performing geometry has been modelled in CST and validated with PIC simulation. 

\begin{figure}[hbt!]
    \centering
    \includegraphics[width=8.0 cm]{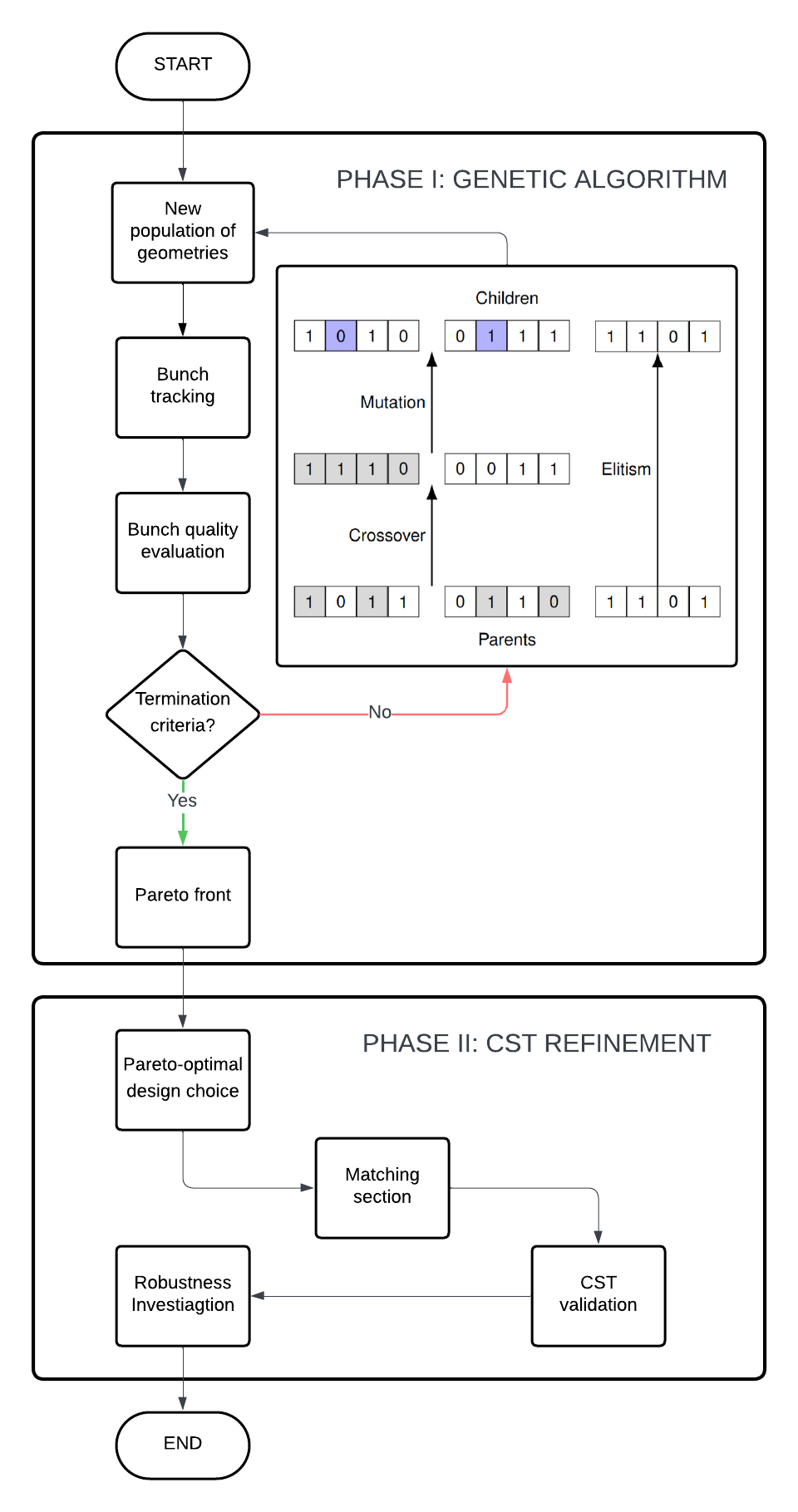}
    \caption{Flow chart of tapered DLW optimisation process. Phase I: Pareto front is approximated through MOGA. Phase II: a single example is chosen and modelled in CST with the addition of a matching section between the first and second segments. Subsequently, solution robustness is assessed by varying the THz pulse amplitude and the bunch-injection timing. }
    \label{fig:cstopt1}
\end{figure}

Importantly, the optimisation outcome is not independent of the initial bunch model. For example, longer initial bunches may result in solutions favouring initial longitudinal compression, while the more transversely spread bunches may benefit from initial transverse compression. In this study, the bunch model used to optimize DLW structures was based on parameters within the reach of 100 keV DC electron guns in the fC-charge regime \cite{DCgun_fC} (Table I).

\section{Bunch quality-preserving geometries}

The result of MOGA optimisation is a Pareto front that, given it generalized the problem well and converged to the global minima of our goal functions, maps all the trade-off solutions of the highest final bunch quality to the corresponding DLW geometries. FIG. \ref{fig:pareto_geometry} shows the final converged objective space. To quantify the trade-offs between the three objectives, we computed the Spearman correlation coefficient $\rho$, where the null hypothesis of no correlation could be rejected (with its probability being $p<0.05$). To demonstrate the evolution of beam dynamics, one example case was selected for each geometry type (Table I). These were chosen using a simple multiplicative metric to be maximized: $CR\cdot\Delta E_{RMS}^{-1}\cdot\epsilon_{xy}^{-1} $. However, any geometry from the final objective front (within a cluster of solutions in FIG. \ref{fig:pareto_geometry}) could be chosen based on trade-off preference.

\subsection{Rectangular DLW}

The solutions within the final objective space share some similar trends. These can be discussed by focusing on a single example corresponding to the desired high-quality final bunch. The genetic algorithm arrived at a specific scheme, illustrated in FIG. \ref{fig:velocity_and_size_emit}, in which at each stage (of constant dielectric thickness) the bunch was overtaken by the accelerating waveform crest ($v_e<v_p$) and then, after some energy (velocity) gain, the accelerating crest was overtaken by the bunch again ($v_e>v_p$). The alternating velocity difference results in alternating longitudinal and transverse focusing/defocusing effects discussed in Section \ref{sec:geom_opt}. This oscillation was both an inevitability of using a limited number of fixed-$v_p$ sections and the phase-alignment that was expected to perform well, as it balances the conflicting requirements of the longitudinal and transverse beam dynamics.

\begin{figure}[hbt!]
    \centering
    \includegraphics[width=8.6 cm]{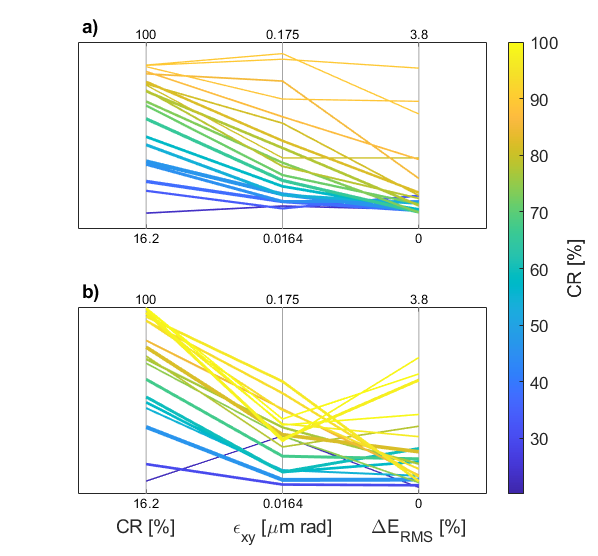}
     \caption{\label{fig:pareto_geometry} Rectangular a) and cylindrical b) DLW plot of final MOGA objective space for non-dominated solutions. Each line marks three bunch qualities of a cluster (of solutions proximate in objective space) of which size corresponds to line width. The capture rate, CR, is defined as in \cite{NixLINAC}.}
\end{figure}

The Pareto-like front depicted in FIG. \ref{fig:pareto_geometry} a) provides insight into the interplay between bunch quality objectives. The solutions were clustered using MATLAB's k-means algorithm to improve the readability of the plot \cite{Pareto_visualisation}. Within that non-dominated solution set, capture rate was very strongly positively correlated with emittance ($\rho=0.96$) and moderately correlated with relative energy spread ($\rho=0.39$), whereas a weak yet significant correlation was observed between emittance and energy spread ($\rho=0.19$, $p=0.011$). This shows that among the non-dominated solutions there is a clear trade-off between the capture rate, determined mainly by the bunch size in phase space, and the bunch quality represented by transverse emittance and energy spread.

\subsection{Cylindrical DLW}
\begin{figure}[hbt!]
    \centering
    \includegraphics[width=9.0 cm]{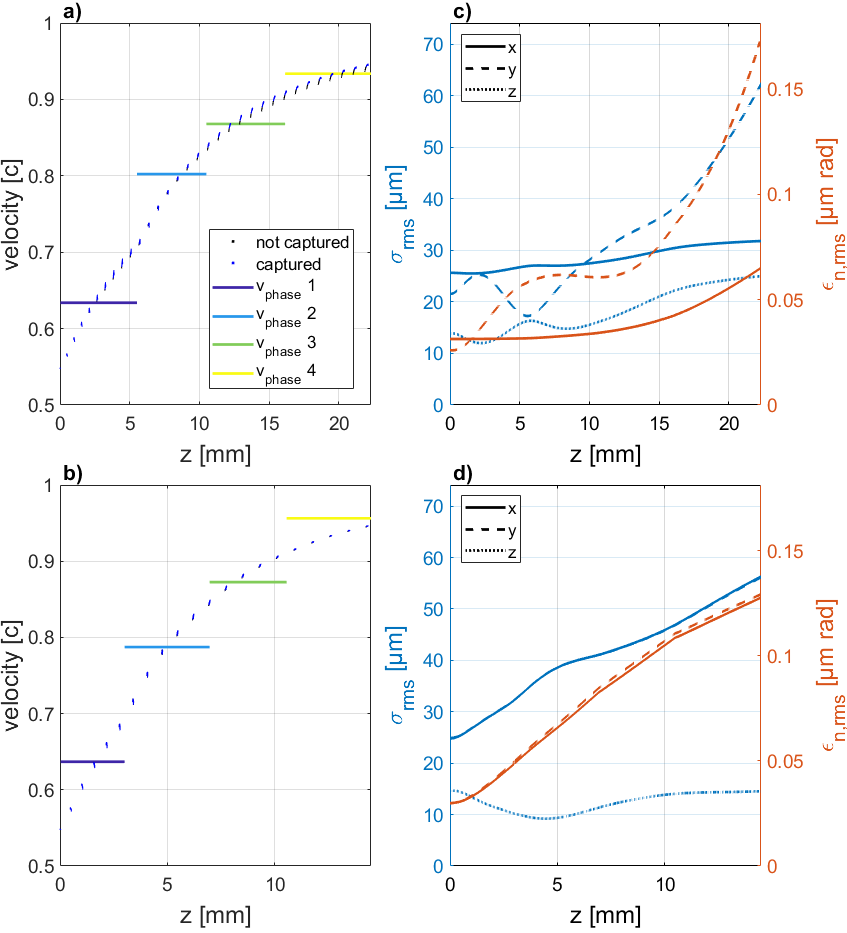}
     \caption{\label{fig:velocity_and_size_emit} Phase velocities (labelled as thickness of segment) and individual particle velocities in rectangular a) and cylindrical b) segmented exemplary structures. Bunch stays synchronous with respect to the accelerating waveform in decreasing-thickness (increasing phase velocity) segments. RMS size and normalized RMS emittance evolution of the bunch along the rectangular c) and cylindrical d) waveguide. }
\end{figure}

In the case of cylindrical geometry, similarly to the rectangular DLW case, the solutions followed a scheme where the bunch was overtaken by the accelerating waveform crest ($v_e<v_p$) and then, after some energy (velocity) gain, the accelerating crest was overtaken by the bunch again ($v_e>v_p$), providing alternating focusing and defocusing forces, as shown in FIG. \ref{fig:velocity_and_size_emit}. This time, however, the exemplar bunch transversely defocused throughout most of the accelerating process in exemplary structure shown in FIG. \ref{fig:velocity_and_size_emit} d). The weaker transversely defocusing fields compared to the rectangular DLW allowed the bunch to remain mostly in the longitudinal compression phase without sacrificing the transverse quality, underscoring the potential of the structure to preserve and accelerate ultrashort bunches.

The optimisation objectives shown in FIG. \ref{fig:pareto_geometry} b) also exhibited pairwise correlations. Similarly to the rectangular case, within that non-dominated solution set, capture rate was positively correlated with emittance ($\rho=0.84$) and moderately correlated with relative energy spread ($\rho=0.53$), yet this time no significant correlation was observed between emittance and energy spread ($\rho=0.11$, $p=0.32$). This again underlines the trade-off relationship between the initial size of the bunch and its quality at the end of the accelerating stage. Moreover, the cylindrical waveguide excelled at mitigating transverse emittance growth, which can be directly linked to lower transverse fields experienced by the bunch, resulting in higher field linearity in the central axis proximity.

Both schemes have been shown here to achieve 100\,keV to 1\,MeV acceleration for the example pulse power we used and both are viable design routes for the development of experimental prototypes. The rectangular option may be machined by cutting from flat sheets of quartz to be glued into a waveguide channel, which likely makes it preferable from a machining standpoint, but both remain of interest for future development.

\setlength{\tabcolsep}{10pt}
\begin{table}[hbt!]
\small
\centering
\begin{tabular}{|l|c|c|}

\hline
\multicolumn{3}{|c|}{\textbf{THz pulse parameters}} \\ 
\hline
frequency & \multicolumn{2}{c|}{0.2 THz} \\
$\tau$ & \multicolumn{2}{c|}{135--150 ps (27--30 cycles)} \\ 
Power & \multicolumn{2}{c|}{3.6 MW} \\ 
\hline
\hline
\multicolumn{3}{|c|}{\textbf{Input beam parameters}} \\ 
\hline
$E$ & \multicolumn{2}{c|}{100.00\,keV} \\ 
$\Delta E_{RMS}$ & \multicolumn{2}{c|}{0.05\,keV} \\ 
$\sigma_z$ & \multicolumn{2}{c|}{15\,$\mu$m (91\,fs)} \\ 
$\sigma_x$, $\sigma_y$ & \multicolumn{2}{c|}{25\,$\mu$m} \\ 
$\epsilon_{xy}$ & \multicolumn{2}{c|}{0.030\,$\mu$m} \\ 
Charge & \multicolumn{2}{c|}{0.50\,fC} \\ 
\hline
\hline
\multicolumn{3}{|c|}{\textbf{Post-acceleration beam parameters}} \\  
\hline
 & \textbf{Rectangular} & \textbf{Cylindrical} \\
\hline
$E$ & 1.0655\,MeV & 1.0805\,MeV \\ 
$\Delta E_{RMS}$ & 4.3 \,keV (0.40$\%$) & 1.8\,keV (0.16$\%$) \\ 
$\sigma_x$, $\sigma_y$ & 31.8\,$\mu$m, 62.5\,$\mu$m & 57.4\,$\mu$m, 57.1\,$\mu$m \\   
$\sigma_z$ & 25.0\,$\mu$m (89\,fs) & 14.8\,$\mu$m (53\,fs) \\ 
$\epsilon_{xy}$ & 0.106\,$\mu$m\,rad & 0.131\,$\mu$m\,rad \\ 
Charge & 0.40\,fC & 0.48\,fC \\  
\hline
\hline
\multicolumn{3}{|c|}{\textbf{DLW geometry}} \\  
\hline
 Segment & \textbf{Rectangular} & \textbf{Cylindrical} \\
\hline
$L_1, d_1$ & 5.50\,mm, 299\,$\mu$m & 2.98\,mm, 376\,$\mu$m \\ 
$L_2, d_2$ & 4.97\,mm, 210\,$\mu$m & 3.96\,mm, 278\,$\mu$m \\ 
$L_3, d_3$ & 5.65\,mm, 196\,$\mu$m & 3.58\,mm, 257\,$\mu$m \\ 
$L_4, d_4$ & 6.16\,mm, 186\,$\mu$m & 3.92\,mm, 243\,$\mu$m \\ 
\hline

\end{tabular}
\caption{Bunch parameters before and after acceleration (simulated in MATLAB) in optimized DLW for two example geometry cases. $L_n$ and $d_n$ represent the length and dielectric lining thickness of the $n-th$ constant-thickness segment.}
\end{table}

\subsection{\label{sec:IV_sensitivity}Sensitivity to manufacturing \& operational deviations}

\begin{figure}[hbt!]
    \centering
    \includegraphics[width=8.6 cm]{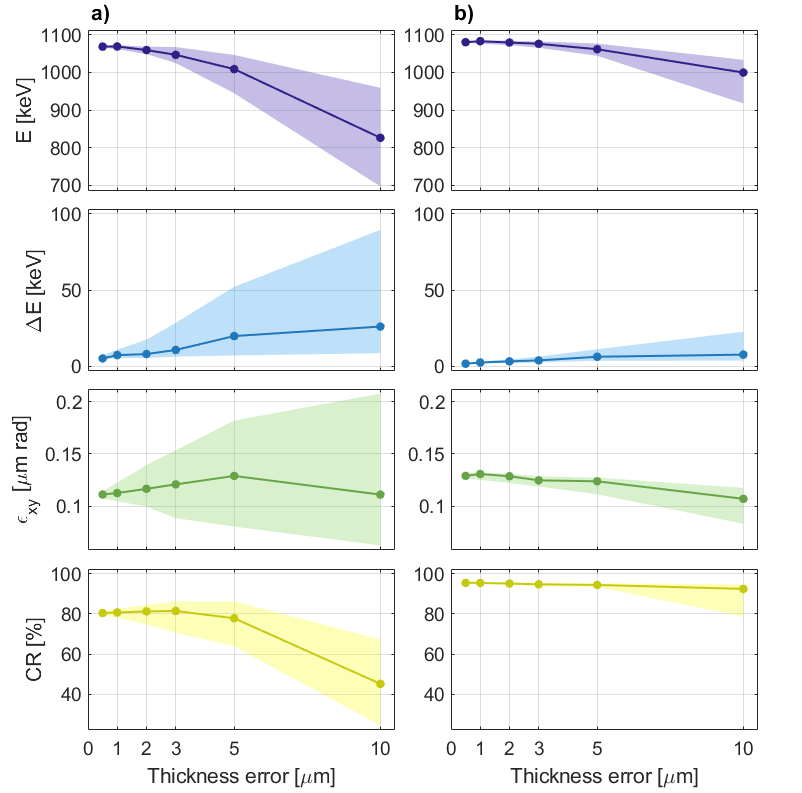}
     \caption{\label{fig:error}Final bunch parameters distributions with various thickness errors applied on  rectangular a) and cylindrical b) DLW (simulated in MATLAB). Each point represents a population of 100 samples. The line corresponds to the median, while the colored region is bounded by $1^{st}$ and $4^{th}$ quartiles.}
\end{figure}

To probe the susceptibility of the final bunch quality to dielectric lining manufacturing tolerance, we varied each of the segments' thicknesses by an error sampled from a random uniform distribution of increasing limit, FIG. \ref{fig:error}. As expected, the higher variance in lining thickness led to degrading final bunch charge and energy. The drop of median normalized emittance for greater random offsets can be explained by the general trend where the uncaptured particles tend to contribute the most to higher emittance due to greater x and y non-linearity of the fields away from the central axis. Moreover, the cylindrical geometry with the same vacuum gap appears more robust to thickness variations, owing to its thicker dielectric lining compared to an equivalent rectangular DLW.

\begin{figure}[hbt!]
    \centering
    \includegraphics[width=8.6 cm]{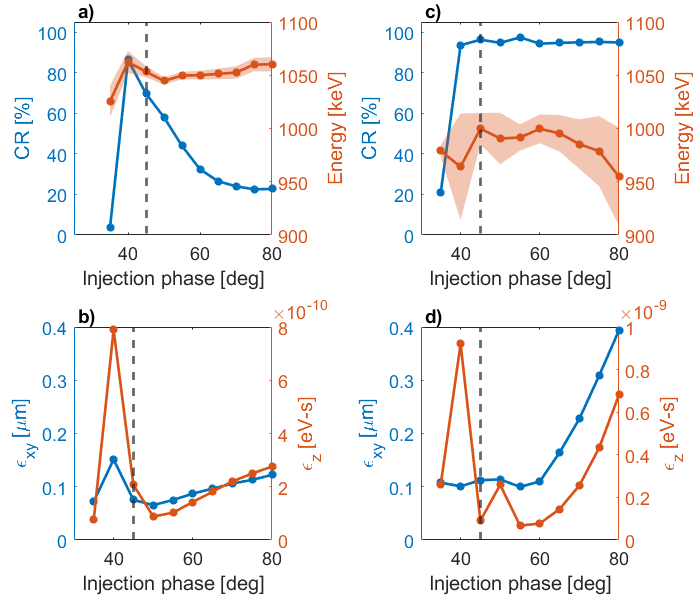}
     \caption{\label{fig:phase_variation_cylindNrect} CST simulation: a) and b) show the change in final bunch parameters in rectangular DLW, while c) and d) in cylindrical DLW, with varying injection phase. The colored regions mark $\pm E_{RMS}$ band. The structures were designed for $+45\degree$ injection (marked with a dashed line) relative to the accelerating crest. The $\epsilon_z$ represents projected RMS longitudinal emittance in $(t, \,\gamma)$ space centred at the bunch.}
\end{figure}

\begin{figure}[hbt!]
    \centering
    \includegraphics[width=8.6 cm]{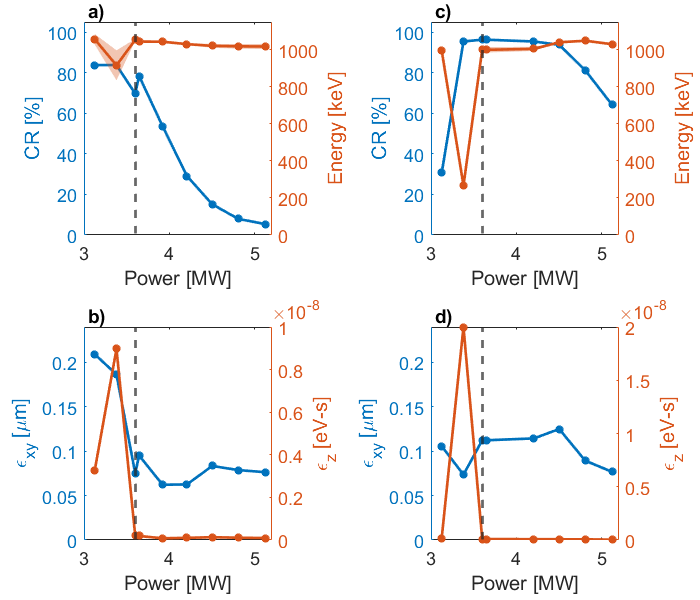}
     \caption{\label{fig:amp_variation_cylindNrect} CST simulation: a) and b) show the change in final bunch parameters, in rectangular DLW, while c) and d) in cylindrical DLW, with varying power excited in DLW. The colored regions mark $\pm E_{RMS}$ region. The structures were designed for $3.6$ MW (marked with a dashed line).}
\end{figure}

The designs produced by the optimisation process were tested by particle-in-cell (PIC) simulation in CST Studio Suite. The analytic model and optimizer were found to accurately produce viable designs, with results that were closely replicated in PIC simulation. Simulations were then used to study the sensitivity to pulse and bunch parameters i.e. errors in injection timing and signal amplitude. FIG. \ref{fig:phase_variation_cylindNrect} shows that in the cylindrical DLW case, the charge capture rate will remain high for the higher but not lower injection phase, where the particles fail to be accelerated past the initial mismatch level in time. The charge capture of the rectangular DLW exhibits sensitivity to both higher and lower injection phases. In cases of varying delivered power (shown in FIG. \ref{fig:amp_variation_cylindNrect}), and thus varying accelerating field strength, the energy remained close to the designed value over a relatively wide range. In the cylindrical case, increasing power did not significantly degrade the bunch parameters over a wide range, while in the rectangular case, stronger fields accelerated bunch clipping, lowering the capture rate.

The analytic field-based simulations have not included space-charge effects, as the accelerated charge was modest $0.5$\,fC. However, an additional simulation including the space-charge interaction was run in MATLAB. The result did not differ significantly for the chosen bunch charge, and for $10$\,fC it diverged only slightly, with a $5\%$ increase in RMS bunch radius and length, and an $8\%$ relative increase in the energy spread in the rectangular case. Similarly, in the cylindrical case, the space charge had no significant effect on the final bunch at $0.5$\,fC, and at $10$\,fC, it caused an $8\%$ increase in the RMS bunch radius and length, and a $5\%$ relative increase in the energy spread. Hence, assuming the higher charge can be delivered within the same given initial phase-space distribution, the structure is able to support acceleration of much higher-charge bunches.

\section{Discussion}

This study focused on a specific goal of accelerating electron bunches from 100\,keV to 1\,MeV. Keeping final energy as one of the optimisation objectives to be maximized (instead of applying penalty) was considered throughout this work. However, this approach led to the spread of final non-dominated solutions across different energy regimes, making the trade-off analysis more challenging. It also resulted in significant variation of excitation pulse length requirements. Nevertheless, the flexibility of optimization technique described allows for freedom in the choice of final bunch energy.

A significantly higher-energy THz pulse could be used to accelerate particles at a higher gradient. That would, however, impose a strict requirement on geometry precision as faster acceleration leads to shorter structures. The shorter structure means the mode needs to adapt more rapidly, increasing the reflections. That would most likely require a closer approximation to adiabatic structure, leading to greater sensitivity to dielectric-thickness error, or could leave insufficient waveguide length for the mode to adapt to the changing geometry, causing significant mode reflection and conversion.

Although the RF guns can operate at similar accelerating gradients, the technique allows for eliminating RF-induced jitter thus delivering intrinsic synchronization with the laser, which is highly desired in ultrafast science applications, such as pump-probe experiments. This may however come at cost of greater amplitude jitter, which reduces its advantage.

There have been a number of publications proposing THz accelerating structure designs at similar energy regimes \cite{cylindDLWtaper, cylindDLW} starting at 200\,keV ($v = 0.70 \,c$) and 400\,keV ($v = 0.83 \,c$), respectively. However,  the difficulty arising from the phase slippage in the travelling-wave accelerator increases drastically with lower energies in this relativistic regime of hundreds of keV, and the techniques presented here overcome it at even lower energy starting point with modest THz pulse energies while mitigating bunch deterioration. Therefore, the next THz-driven acceleration stages operating in a relativistic regime can sustain bunch-mode synchronization at much higher gradients.

\section{Conclusions}
In this work, the MeV-level acceleration of a 100\,keV electron bunch in THz-driven tapered DLW was demonstrated. By precisely tailoring the waveguide geometry, the electromagnetic field profile and phase/group velocities are tuned to maintain synchronization between the beam and the accelerating mode over extended distances before a phase walk-off occurs. The THz-regime DLW optimisation process was presented here based on the analytic model of the accelerating mode in DLW, benchmarked against equivalent CST Studio Suite PIC solver results. The model was implemented in a particle tracker to simulate the beam dynamics of an externally injected 100\,keV electron beam for various DLW geometries. A genetic algorithm was employed to identify the optimal geometry, based on the objective function that accounts for final bunch qualities such as emittance, energy spread, and charge. Simulations indicate that low-emittance 0.5\,fC bunches can be accelerated from 100\,keV to over 1\,MeV in approximately 20-mm-structure with 0.5\,mJ of energy in a multi-cycle 0.2\,THz pulse. The advantage of our design method choice is its geometric simplicity, which falls well within modern manufacturing capabilities, and the insight into the trade-off between the parameters representing the quality of the final bunch.

We have demonstrated viable solutions in both the rectangular and cylindrical geometry regimes for dielectric-lined waveguides. The former is less complex from a manufacturing perspective and can offer stronger focusing fields, while the latter is more robust to deviations from designed parameters, offers stronger on-axis field for a given modal power and excels at preserving transverse emittance of the electron beam. Both choices showcase a potential for tabletop structures providing MeV-energy electron beams for ultrafast science experiments or subsequent synchronous accelerating stages.
\vspace{\baselineskip}
\section{\label{sec:ack}Acknowledgments}
This work was supported by the United Kingdom Science and Technology Facilities Council [Grant No. ST/V001612/1].

\bibliography{refs}% Produces the bibliography via BibTeX.

\end{document}